\documentclass{pramana}

\usepackage{graphicx,amsmath,bm}

\usepackage{caption}
\usepackage{subcaption}
\begin{document}

%%paper title
%%For line breaks \\ can be used within title 
\title{Dynamics of slow and fast systems on complex networks}

%%author names are separated by comma (,) 
%%use \and before the last author name 
%%\textsuperscript{number} is used for affiliation
%%use a * along with the number separated by comma
%% for the  author for correspondence

\author{Kajari Gupta, G. Ambika\textsuperscript{*}}
\affilOne{ Indian Institute of Science Education and Research, Pune\\}

%%escape two column mode for title, affiliation and abstract
%%by giving \twocolumn command as shown

\twocolumn[{

\maketitle

%%include \corres to print the corresponding author Email id
\corres{g.ambika@iiserpune.ac.in}

%%include \msinfo for
%%manuscript information such as
%%received, revised and accepted dates
%%
%\msinfo{1 January 2015}{1 January 2015}{1 January 2015}

%%abstract
\begin{abstract}
We study the occurrence of frequency synchronised states with tunable emergent frequencies in a network of connected systems. This is achieved by the interplay between time scales of nonlinear dynamical systems connected to form a network, where out of N systems, m evolve on a slower time scale. In such systems, in addition to frequency synchronised states, we also observe amplitude death, synchronised clusters and multifrequency states.  We report an interesting cross over behaviour from fast to slow collective dynamics as the number of slow systems m increases.  The transition to amplitude death is analysed in detail for minimal network configurations of 3 and 4 systems which actually form possible motifs of the full network.
\end{abstract}

%%insert keywords separated by comma using \keywords{words}

%%include \pacs{number} to print the PACS number
%\pacs{12.60.Jv; 12.10.Dm; 98.80.Cq; 11.30.Hv}

}]
%%close the twocolumn escape here

%%include \doinum{number}for the DOI number in the header
%%include \volnum{number} for the volume number in the header
%%include \year{yyyy} for  year of publication in the header
%%include \pgrange{num--num} page range of article in the header
%%include \artcitid{num} for the article citation id
%%include \lp to print last page of the article
%%include \setcounter{page}{pagenum} for the exact starting page of the article

%\doinum{12.3456/s78910-011-012-3}
%\artcitid{\#\#\#\#}
%\volnum{123}
\year{2017}
\pgrange{1--6}
\setcounter{page}{1}
\lp{6}

\section{Introduction}
Many complex systems that occur in physical, biological, chemical and geophysical contexts have large number of sub systems or units interacting with each other. The collective behavior of such highly interconnected dynamical systems is being studied recently using the framework of complex networks with the sub units representing the nodes and the connectivity among them indicated by links. The varied levels of their complexity arise from the nonlinearity of the nodal dynamics, nonlinear or stochastic processes that happen on them and the nature of the interactions reflected in the complex topology of the networks. Some of the well studied emergent phenomena in such contexts are synchronization \cite{alex08}, amplitude death\cite{resmi12}, clustered states and chimera states\cite{{yun14},{scholl16}}.

Recent studies have emphasized the role of heterogeneity in such real world complex systems.  Heterogeneity can arise in complex networks in various ways, such as degree heterogeneity, heterogeneity in the weights or nature of coupling, parameter mismatch and dissimilar nodal dynamics, dynamical or time varying couplings etc. This can affect processes happening on the network like rumour or epidemics spreading\cite{{rui14},{ce16}} ecosystem stability in ecological networks\cite{{wen14},{sha14}} and plays an important role in the diversity and organizations of social network\cite{law92} or breaking synchrony in oscillator network \cite{den04}. In power grids and internet it has been shown that heterogeneity due to different loads on nodes can cause a cascaded effect of overload failures\cite{adi02}. 

In this work, we present the importance of heterogeneity caused by different time scales of the nodal dynamics on the network. In this context we note that dynamical processes containing different time scales occur in systems such as modulated lasers and chemical reactions\cite{{das13},{zunino11}}. In biological systems, dynamics with time scale of days interacts with dynamics of biochemical reactions with sub seconds time scale\cite{{mit02}}. In neuronal network, electric signals of neurons get directly or indirectly affected by intercellular processes of various time scales\cite{kay03}. In the context of weather and climate systems of earth, one sees sub systems of widely varying time scales gets strongly coupled to each other like tropical atmospheric ocean systems\cite{{dav91},{lind08}}. It is reported that climate sensitivity and stabilization occur through process with a range of different time scales\cite{{will08},{mee01}}. There are also studies on populations with time scale diversity\cite{sil03} and spatio-temporal chaos with cascade of bifurcations caused by interactions among different time scales\cite{fuji03}. 

Recently, a basic study on coupled nonlinear systems with fast and slow time scales, reported interesting emergent states like amplitude death state and frequency synchronized states\cite{kaj16}. However a detailed analysis of nodal dynamics with different time scales has not been done in the context of complex networks. We find there are many interesting open questions to be addressed regarding the possible emergent dynamics, its characterization and transitions when nodal systems follow multiple time scales.     

In this study, we consider the role of heterogeneity in the emergent collective behavior in a network of nonlinear dynamical systems, where the heterogeneity arises only from the difference in the time scales of nodal dynamics. To make this a specific feature and bring out effects of time scale mismatch of connected systems, we consider an otherwise homogeneous network topology of fully connected network where out of N systems m evolve at a slow time scale compared to others. We observe dynamical states like synchronized clusters, multi frequency states, phase synchronized states and phenomena like amplitude death and cross over behavior in the collective dynamics that occur in the network as m is varied for different values of mismatch of time scales and coupling strength.

\section{ Network of slow and fast systems}
We consider a network of N identical systems in which m evolve on a slower time scale. This subset of oscillators with lower timescale is defined as S.  The equations of n dimensional systems governing the dynamics are given by
\begin{eqnarray}
\dot{X_i} = \tau_i F(X_i)+G \epsilon\tau_i\displaystyle\sum_{j=1}^N A_{ij}(X_j-X_i)
\label{neteqn}
\end{eqnarray}
where $ \tau_i = \tau $ if $ i $ $\in$ S, $ \tau_i = 1 $ otherwise. G is an n x n matrix which decides which variables are to be coupled. Here we take G = diag(1, 0, 0 ..) which means x variable of the $ i^{th}$ oscillator is coupled diffusively with the x variable of $j^{th}$ oscillator. $ A_{ij} $ is adjacency matrix of the network defining its topology or connectivity. Since we consider a fully connected network, $ A_{ij}=1 $ for all i and j except i=j. In this study we consider the dynamics at each node as a periodic R{\"o}ssler system given by

% We study the case of system size N=100 and analyse how the slowness of m of the systems can affect the dynamics of the whole network.

\begin{eqnarray}
 \dot{x}_i&=& \tau_i(-y_i-z_i) +\tau_i \epsilon \displaystyle\sum_{j=1}^N A_{ij}(x_j-x_i)
\nonumber\\
\dot{y}_i&=&\tau_i(x_i+ay_i) 
 \nonumber\\
\dot{z}_i &=& \tau_i(b+z_i(x_i-c))
\label{prosseqn}
\end{eqnarray}
The parameters are chosen as a=0.1, b=0.1 and c=4 so that dynamics is in the periodic region.

One of the main results of our study is the suppression of dynamics in the whole network, or amplitude death (AD) that occurs for sufficient timescale mismatch and coupling strength for a range of values of the number of slow systems, m. To identify this range, we calculate the difference between global maxima and global minima ($A_{diff}$) in the x-time series of each system. The average of this difference is plotted for different m values in a network of 100 systems for in Fig~\ref{para}. The region for which $<A_{diff}>=0$, corresponds to suppression of dynamics and the region of synchronized fixed point or amplitude death. 
\begin{figure}[H]
\centering
\includegraphics [width = 0.6\columnwidth]{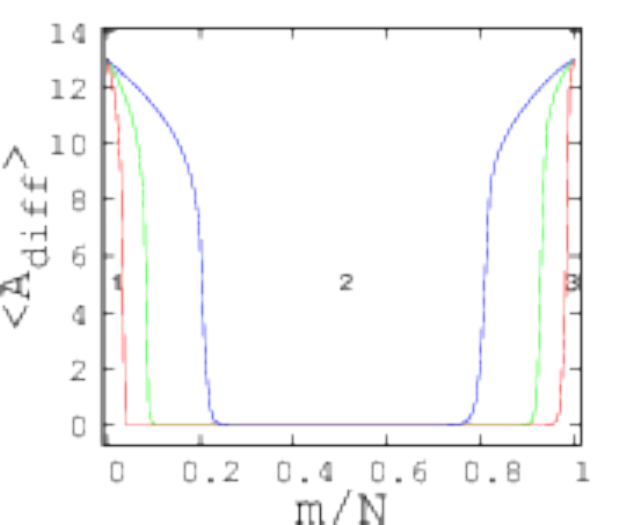}
\caption{\label{trans}(colour online)Average amplitude difference $ <A_{diff}> $ of N coupled R{\"o}ssler systems plotted with the fraction, m/N of slow systems for $N=100$ keeping $\epsilon=0.03$ and varying $\tau=0.1$(red),$0.35$(green),$0.5$(blue) . $ <A>=0 $ corresponds to AD indicated as region 2. }
\end{figure}
We now present in detail, the nature of dynamics in all the three regions marked in Fig~\ref{trans}. 
\subsection{Synchronized clusters, multi-frequency states and frequency synchronization for small m}
\begin{figure}[H]
\centering
\includegraphics [width = \columnwidth]{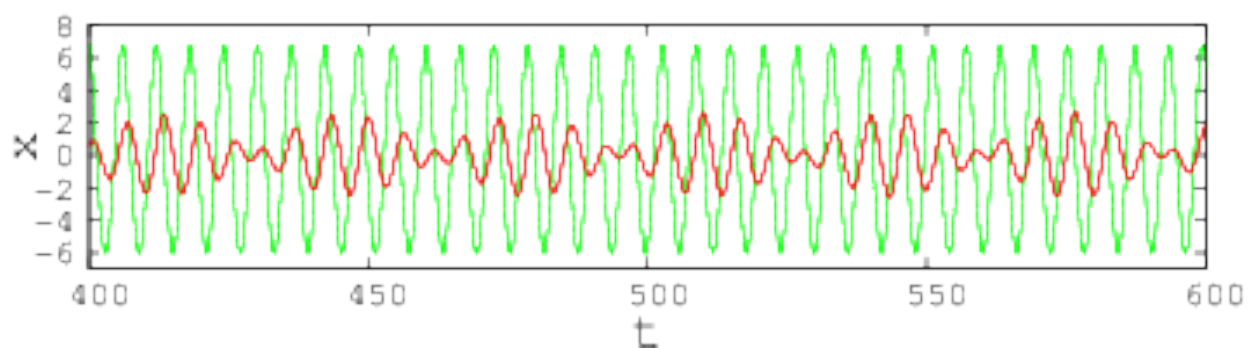}
\caption{\label{twofreq}  Time series of x variable are plotted for 3 slow and 3 fast systems out of 100 total number of systems. This plot shows two frequency state for slow(red) systems while fast(green) systems remain periodic for $ \tau=0.8 $, $ \epsilon=0.001, m=10 $.}
\end{figure}

In the region 1 of Fig~\ref{trans}, where number of slow systems m is small (say m$\approx$10), under weak coupling ($\epsilon \approx$ 0.001) and large $ \tau $, we observe the whole network will split into two separate synchronized clusters. The dynamics of one cluster of slow systems is synchronized two-frequency state of small amplitude oscillations and the fast systems are synchronized among themselves into periodic oscillations having larger amplitudes. This behavior is shown in Fig.~\ref{twofreq} where the asymptotic time series of three typical systems from the fast set and slow set are shown.

\begin{figure}[H]
\centering
\includegraphics [width = \columnwidth]{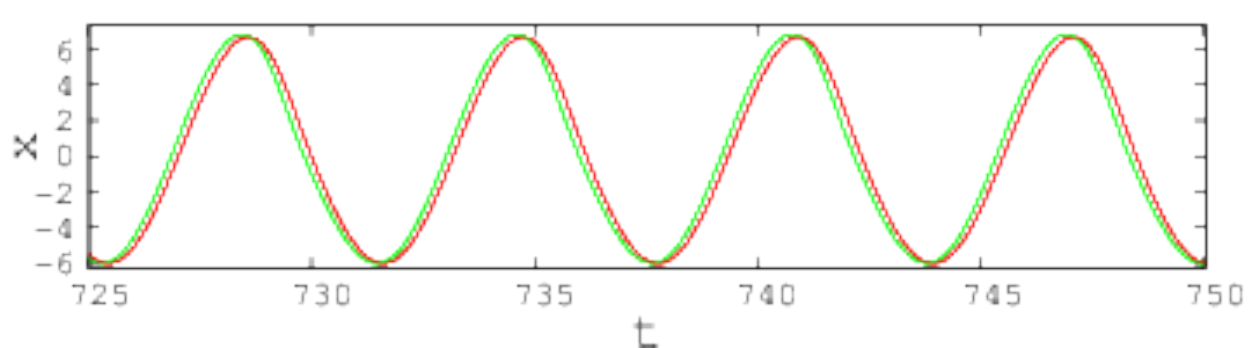}
\caption{\label{net}(colour online)Synchronized clusters of slow and fast dynamics in the network for $ \tau=0.8 $, $ \epsilon=0.03 $. Here the time series of the x variable are plotted for 3 typical slow(red) and 3 fast(green) systems. The two clusters are frequency synchronized with a phase shift between them. }
\end{figure}

As the coupling strength increases, the two synchronized clusters of fast and slow systems get into frequency synchronized states. This is indicated by time series in Fig.~\ref{net} for $\tau=0.8 $, $ \epsilon=0.03 $. Here the frequency synchronized clusters are separated by a phase shift. Here also, the amplitude of the cluster of intrinsic fast systems is more than that of the slow one. When the timescale mismatch increases the behavior remains the same with increasing difference in amplitudes between fast and slow systems.

\subsection{Suppression of dynamics and frequency synchronization for moderate m}
\begin{figure}[H]
\centering
\includegraphics [width = \columnwidth]{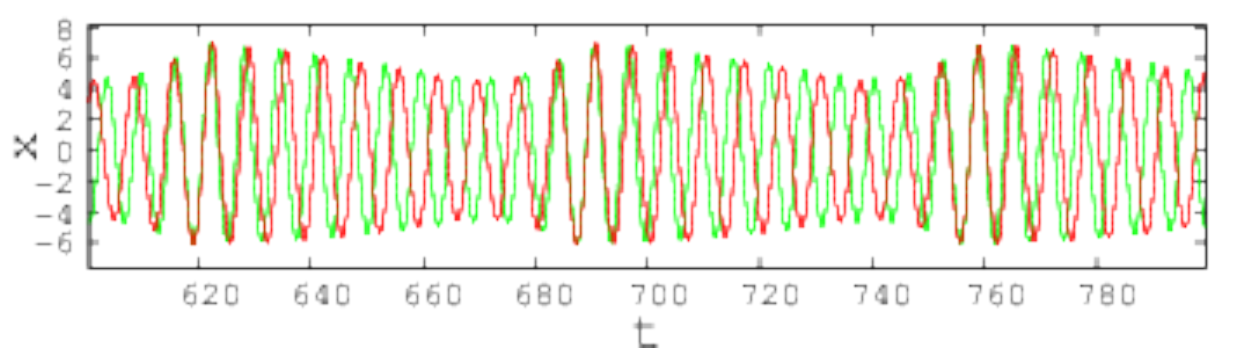}
\caption{\label{ratio}(colour online) Time series of x variable are plotted for 3 slow and 3 fast systems out of 100 systems in the network. This shows two frequency states of slow(red) and fast(green) systems synchronized within the clusters for m=50,$\tau=0.9,\epsilon=0.001$}
\end{figure}

As the number of slow systems m increases, corresponding to region 2 in Fig~\ref{trans}, for weak coupling, the systems go into two frequency states as shown in Fig~\ref{ratio}. In this case also slow systems and fast systems are synchronized within each set but both having comparable amplitude. However, the two clusters have matching frequencies for the smaller of the two emergent frequencies while their larger frequencies differ. 

\begin{figure}[H]
\centering

\includegraphics [width = \columnwidth]{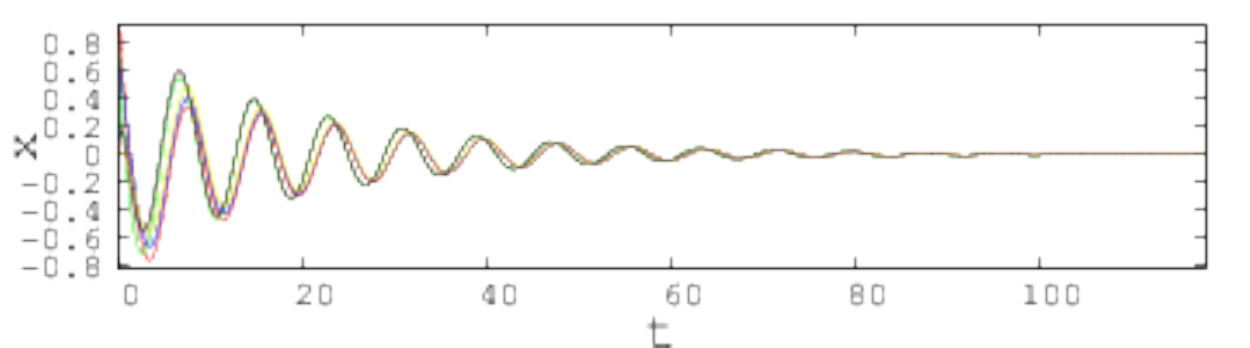}
\caption{\label{ad}(colour online)Amplitude death state for fully connected network of slow and fast systems for $ \tau=0.4 $, $ \epsilon=0.03, m=50 $. Here the time series of x variable of 3 fast and 3 slow systems are plotted.}
\end{figure}

\begin{figure}[H]
\centering
\includegraphics [width = 0.7\columnwidth]{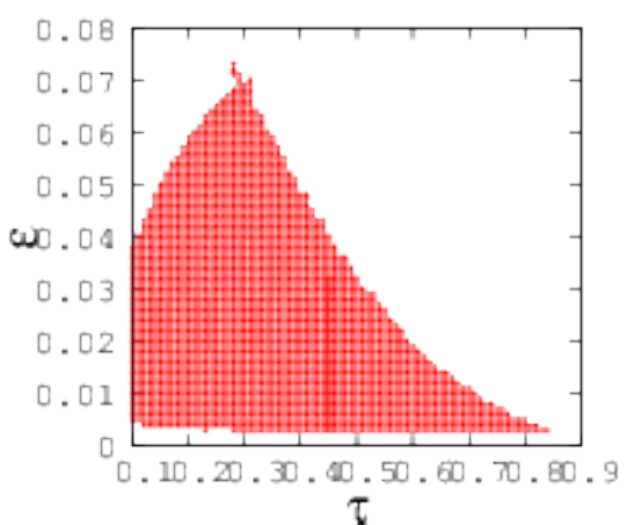}
\caption{\label{para}(colour online)Region of amplitude death in ($\tau,\epsilon$) plane for m=50.}
\end{figure}

In this case, under strong coupling, mismatch in time scale of interacting slow and fast systems causes the suppression of dynamics in the whole network, observed for the specific range of $\tau$ and $\epsilon$. This state of amplitude death(AD) is shown in Fig~\ref{ad}.
 As shown in Fig.~\ref{trans}, AD can occur for a specific range of m values. For any value in this range, we can isolate the region of AD in ($\tau,\epsilon$) parameter plane. For this we numerically calculate the average difference between global maxima and global minima ($A_{diff}$)of all systems in the network and mark the region( in red) where this is zero as shown in Fig~\ref{para}.   

In the regions outside the AD region, when mismatch in timescale is low and $\epsilon$ is high, we once again observe the systems separating into two clusters, where fast systems belong to one cluster and slow systems belong to another, while being frequency synchronized with each other with a phase shift.

\begin{figure}[H]
\centering

\includegraphics [width = 0.5\columnwidth]{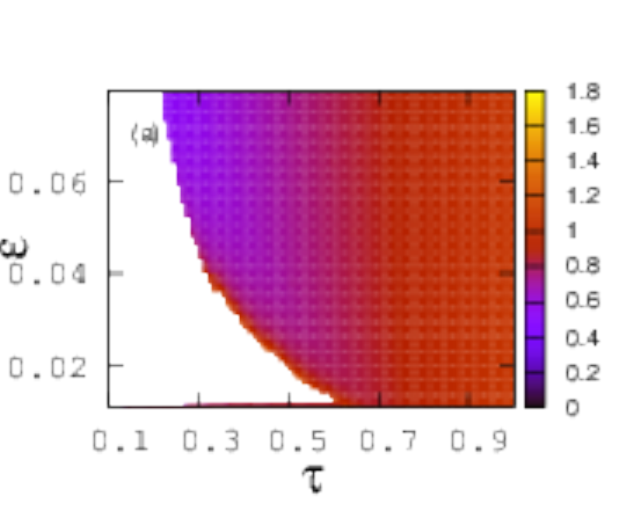}\includegraphics [width = 0.5\columnwidth]{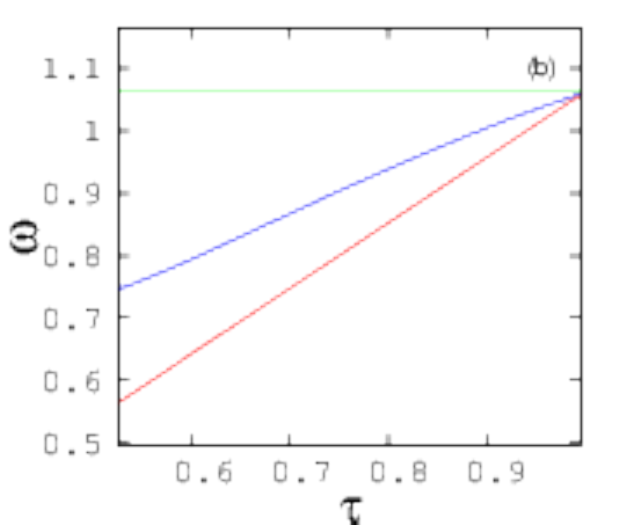}
\caption{\label{ratio2}(colour online)a) Variation of the emergent frequency in ($\tau,\epsilon$) plane for m=50 in frequency synchronized state. The color code is as per the frequency of the emergent state. b) Variation of intrinsic frequency of fast systems (red), that of slow systems(blue) compared with the emergent frequency (green) as $\tau$ is varied for $\epsilon=0.03, m=50$}
\end{figure}

We calculate the frequency of this emergent state from the time series, as reported in the earlier work\cite{kaj16} for parameters in the region outside of amplitude death in the ($\tau,\epsilon$) plane (shown in Fig.~\ref{ratio2}a). The emergent frequencies are compared with the intrinsic frequency of each oscillator in Fig~\ref {ratio2}b.

For the choice of parameters considered in this section, the clusters of fast systems and slow systems evolve with different amplitudes while being synchronized within the cluster. 

\subsection{Cross over behaviour in dynamics for large m}
 As m increases further, in region 3 of Fig~\ref{trans}, the network regains the dynamics from the amplitude death state but follows the slow time scale, in general. When the mismatch is not large or for large $\tau$, the systems form clustered states with slow systems having larger amplitude than fast systems. Thus there is a clear cross behavior in the collective dynamics of the network as m varies. This can be traced in two ways, by the emergent frequency of the frequency synchronized state and by the average amplitude of the synchronized clusters.

\begin{figure}[H]
\centering
\includegraphics [width = 0.5\columnwidth]{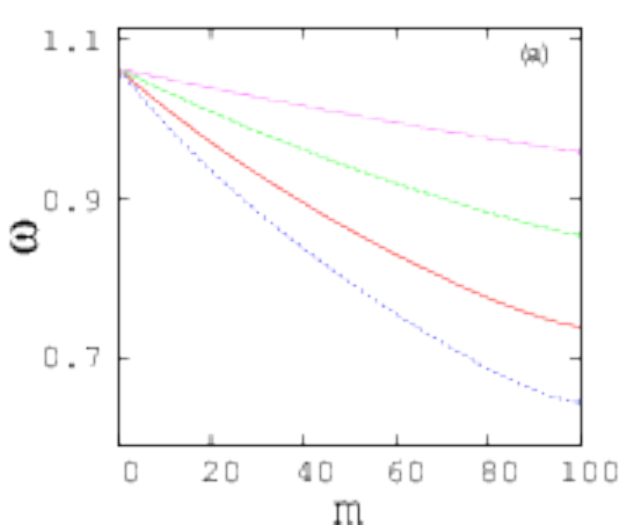}\includegraphics [width = 0.5\columnwidth]{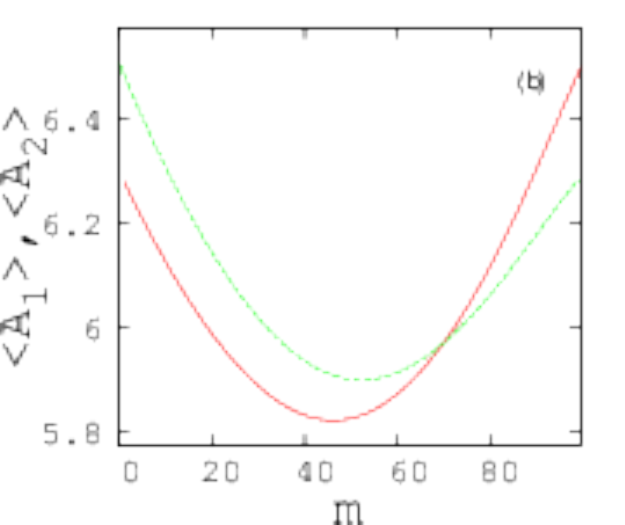}
\caption{\label{ratio3}(colour online)a)Variation of emergent frequency with m for $\epsilon=0.03$ and $\tau$ varied as $0.6,0.7,0.8,0.9$ (from below to above respectively) and b) average amplitude of slow (red) and fast (green) oscillators with m showing crossover behavior for $\epsilon=0.03, \tau=0.7$ at $m_2=70$. }
\end{figure}

We observe that the emergent frequency in frequency synchronized state decreases with the increase of m, as shown in Fig.~\ref{ratio3}a for a particular value of $\tau$ and $\epsilon$. In this case we note that for $\epsilon=0.03$ the frequency becomes less than the average of intrinsic fast and slow frequencies at a critical value of m, say $m_1$, which we study for $\tau=0.6,0.7,0.8,0.9$. Thus this this value of $m_1$ gives the critical value where frequency suppression starts. 

\begin{figure}[H]
\centering
\includegraphics [width = 0.6\columnwidth]{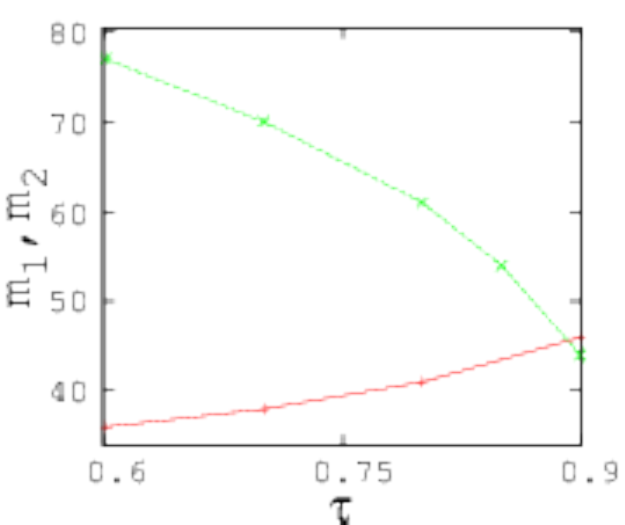}
\caption{\label{cross}(colour online)Variation of crossover thresholds with $\tau$ for $\epsilon=0.03$. Red line denotes the critical number of slow systems ($m_1$) at which frequency suppression starts and green line denotes the same ($m_2$) for amplitude crossover.}
\end{figure}

We also observe another cross over behavior in the amplitudes of oscillations of slow and fast systems when m is increased. To study this we plot the average amplitudes of the fast cluster and slow cluster for $\tau=0.7$ and $\epsilon=0.03$ in Fig~\ref{ratio3}.  We notice that the cross over from a state where the amplitudes of fast systems are larger to one where that of slow systems are larger, happens at m=70. 
We find both these cross over points varies with the parameters $\tau$ and $\epsilon$. To illustrate this we plot $m_1$ and $m_2$ values for cross over behavior, with different $\tau$ values for $\epsilon=0.03$(Fig~\ref{cross}).
\section{Suppression of dynamics in minimal network with differing time scales}
\begin{figure}[H]
\centering
\includegraphics [width = 0.48\columnwidth]{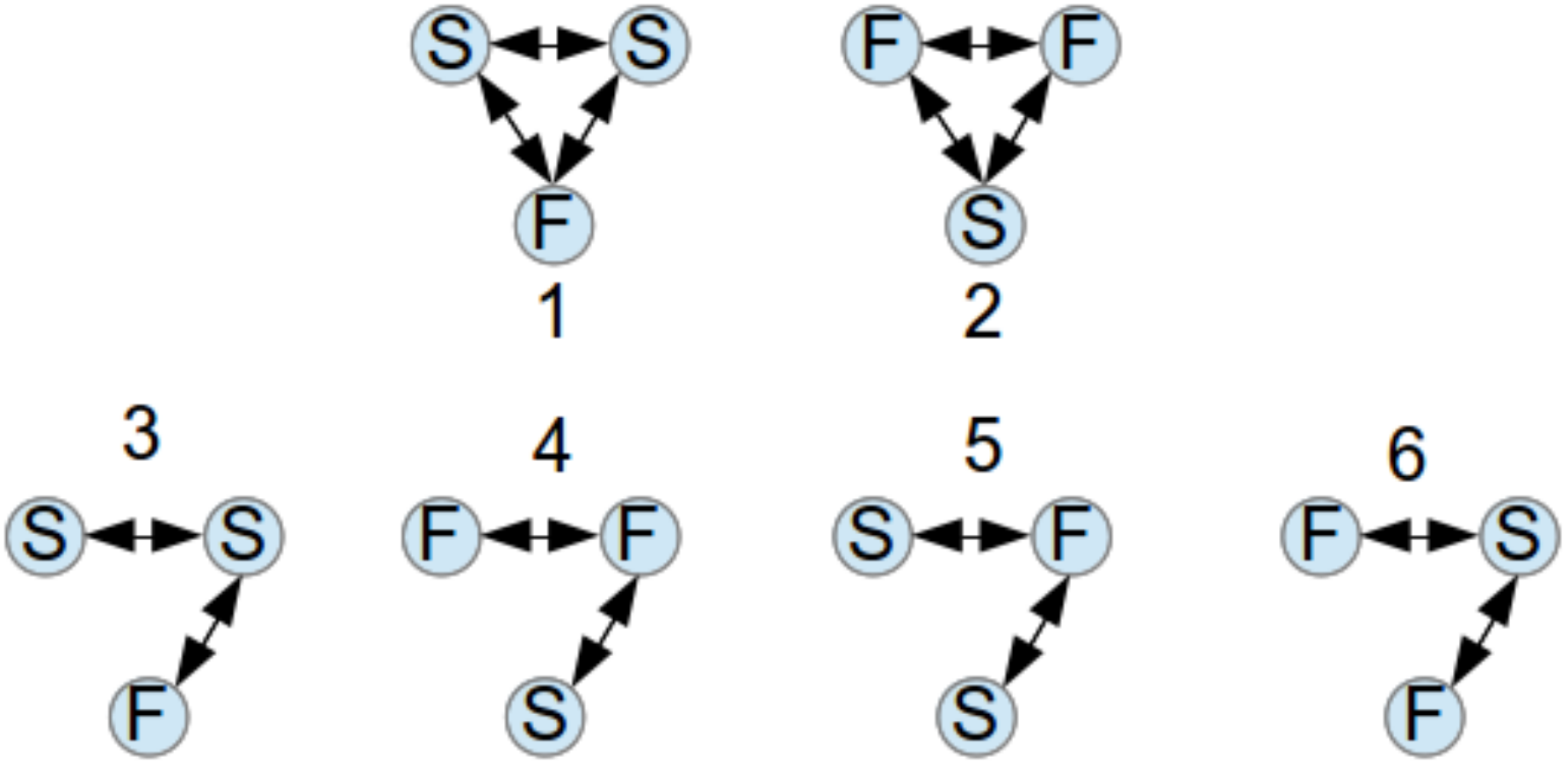}\,\,       \includegraphics [width = 0.48\columnwidth]{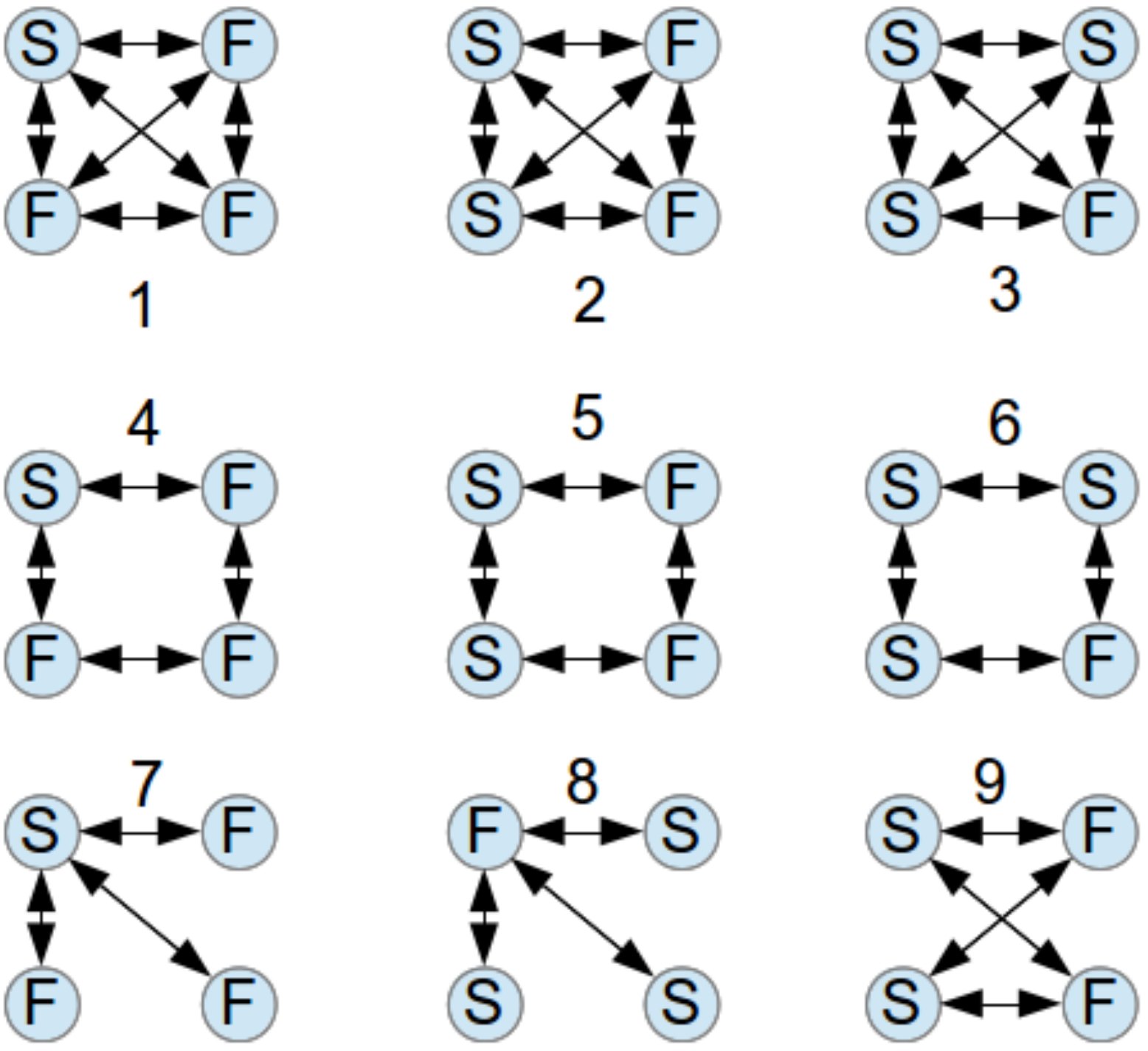}	

\caption{\label{pr_trans}Configurations of minimal networks with 3 systems and 4 systems having different possible m.}
\end{figure}

In this section, we analyze in detail the onset of amplitude death due to difference in time scales of connected systems by considering two minimal configurations of the network with 3 and 4 systems each. We consider in Fig~\ref{pr_trans}, all possible configurations of N=3 and N=4 with two different time scales marked as S-slow and F-fast.  As we can see for system size $N=3$, total number of unique configurations are 6, where m is varied as 1 and 2. In the case of 4 systems, we take configurations of ring and bipartite structure in addition to the fully connected structure with m= 1,2 and 3. These serve as possible motifs in the large network and can be subjected to analytical study for the transition to AD.

The AD in this context corresponds to synchronized fixed point of the system and hence its onset and occurrence can be derived from the stability analysis of the fixed point of the system $(x^*,y^*,z^*)$. For this we calculate the eigenvalues of Jacobian of each configuration, around the fixed point $(x^*,y^*,z^*)$ of the coupled systems for different values of $ \tau $ and $ \epsilon $ and all eigenvalues having negative real parts corresponds to stable region of AD.
The generic form of Jacobian of any network of slow and fast systems can be written as below
\begin{equation}
{\bf J}= ({\bf \tau}{ \bf .}{\bf I}){\bf x}{\bf F}+({\bf \tau}{\bf .}{\bf A}){\bf x}{\bf H}
\end{equation}

where ${\bf \tau}$ is an $NxN$ matrix in which $\tau_{ij}=\tau_i$ equation~(\ref{neteqn}). {\bf I} is NxN identity matrix. Dot product ({\bf .}) is defined by the element wise product of two matrices, and cross product ({\bf x}) is defined as each element of the former matrix being multiplied by the later matrix as a block. {\bf A} is the adjacency matrix. In the case of R{\"o}ssler systems with coupling function as given in equation~\ref{prosseqn}, 
\begin{equation}
 {\bf F} =   \left (\begin{array}{ccc}
-n\epsilon & -1 &  -1  \\
1 & a & 0  \\

z* & 0  & (x*-c) \end{array} \right), \nonumber
  {\bf H} =   \left (\begin{array}{ccc}
\epsilon & 0 & 0  \\
0 & 0 & 0  \\

0 & 0  & 0 \end{array} \right) , \nonumber \\\\
\end{equation}
\begin{equation}
 {\bf 0} =   \left (\begin{array}{ccc}
0 & 0 & 0  \\
0 & 0 & 0  \\

0 & 0  & 0 \end{array} \right)
\label{jacobian1}
\end{equation}

Here, 
$ (x^*,y^*,z^*) = (\frac {c-\sqrt{c^2-4ab}}{2},\frac {-c+\sqrt{c^2-4ab}}{2a},\frac {c-\sqrt{c^2-4ab}}{2a}) $.\\\\
For example in the case of N=4, the Jacobian of the configuration where m=2, and only slow systems are connected to fast systems, shown in (9) in Fig.~\ref{pr_trans}, can be written as below.
\begin{equation}
 J =   \left (\begin{array}{cccc}
\tau{\bf F} & {\bf 0} & \tau {\bf H} & \tau {\bf H} \\
{\bf 0} & \tau{\bf F} & \tau {\bf H} & \tau {\bf H} \\

{\bf H} & {\bf H}  & {\bf F} & {\bf 0}  \\
{\bf H} & {\bf H}  & {\bf 0} & {\bf F}  \end{array} \right) 
\label{jacobian}
\end{equation} 
Here $n=2$ in the matrix {\bf F} as the adjacency matrix {\bf A} has 2 entries of 1 in each row.

\begin{figure}[H]
\centering
\includegraphics [width = 0.5\columnwidth]{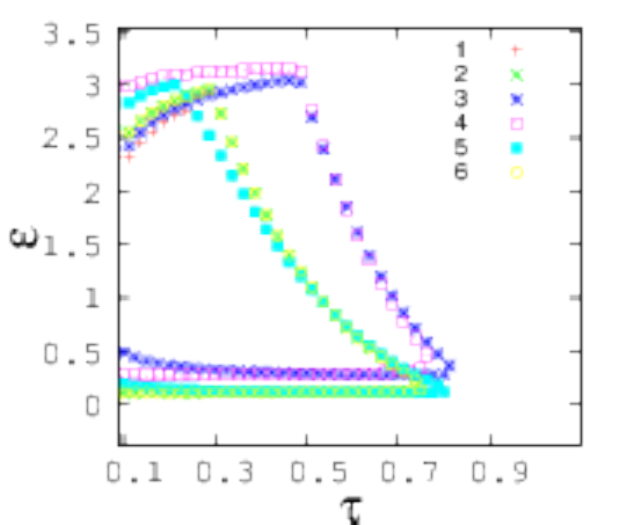}\includegraphics [width = 0.5\columnwidth]{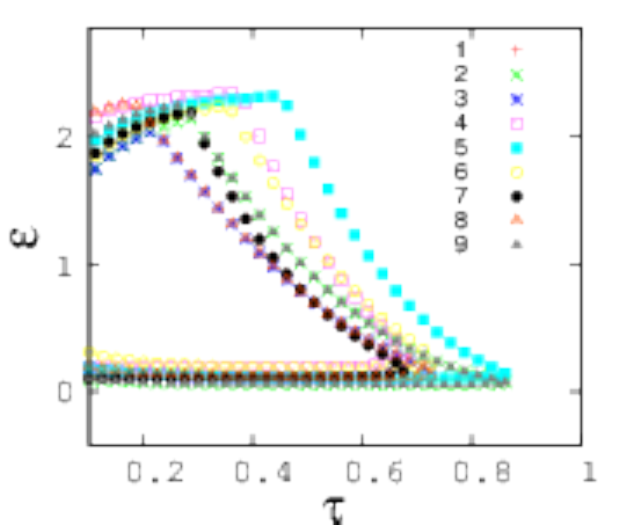}	

\caption{\label{pr_trans1}Transition curves to AD for different configurations or motifs with $N=3$ and $4$}
\end{figure}

We now estimate the eigenvalues of the Jacobian for each configuration for different values of $ \tau $ and $ \epsilon $ and identify the boundary of transition to AD as the point where at least one of the eigenvalues of the corresponding Jacobian goes from negative to positive.  This is repeated for  all the configurations of $N=3$ and $N=4$, and the boundaries of transition obtained in the parameter when in $\tau,\epsilon$ plane are shown in the Fig~\ref{pr_trans1}.
For configurations 1 and 2 in $N=3$ and configurations 1,2 and 3 in $N=4$ shown in Fig.~\ref{pr_trans}  , as expected, the behavior outside AD region is similar to the behaviours described in sec. 2 showing two frequency states, clustering, frequency synchronization.  

\section{Conclusion}
Our study shows how mismatch in time scales of different interacting units affect the emergent collective dynamics of the whole system. In an all to all connected network of N systems with m of them having dynamics of slower time scale, in general, the systems can separate into two clusters, one of slow systems and another of fast systems, each having synchronized dynamics among its units. 

The dynamics of each cluster depends on the strength of coupling, $\epsilon$ and the time scale mismatch $\tau$. The cases where the two clusters settle to frequency synchronization between them and that for which the whole network stabilizes to amplitude death are studied in particular and the transitions between them identified. 

We find an interesting cross over behavior in both frequency of the emergent state as well as in the emergent dynamics as the number of slow systems increases. The cross over points in both cases are studied for different values of time scale mismatch. By taking all possible configurations of minimal networks of size 3 and 4, we analytically study the transition to AD in them. These can serve as substructures or motifs in much larger networks under similar situations. 

It is interesting to note that occurrence of the various phenomena observed in the study, like frequency synchronized clusters, frequency suppressed states, amplitude death etc can be controlled by tuning the parameters m, $\epsilon$ and $\tau$. This means one can 
achieve desired frequency in the output of the whole network by a suitable choice of these parameters. This can be a possible natural process that happens in real world networks like neurons and can have engineering applications in power transmission networks, traffic systems and sensor networks where controlled frequencies play a relevant role.

%%Appendix

\section*{Acknowledgement}

One of the authors (K.G) would like to thank University Grants Commission, New Delhi, India for financial support.

%%use \balance somewhere in the left column of the last page to balance the two columns in the end page

%%References section

\end{document}